\documentclass{aa}
\usepackage{graphicx,txfonts,natbib}
\bibpunct{(}{)}{;}{a}{}{,}
\begin{document}

\title{The dynamics of the nebula \object{M1-67} around the run-away Wolf-Rayet star \object{WR 124}}
\author{M.V. van der Sluys \inst{1}
        \and  H.J.G.L.M. Lamers \inst{1,2}
       }
\titlerunning{The dynamics of the nebula M1-67 around the star WR~124}
\authorrunning{Van der Sluys \& Lamers}
\offprints {H.J.G.L.M. Lamers, \email{lamers@astro.uu.nl}}

\institute{ Astronomical Institute, Princetonplein 5, NL-3584 CC Utrecht,
            the Netherlands,
            {\tt(sluys@astro.uu.nl)} and
            {\tt(lamers@astro.uu.nl)}
            \and
            SRON Laboratory for Space Research, Sorbonnelaan 2, NL-3584 CA
            Utrecht, the Netherlands
          }

\date{Received July 18, 2002 / Accepted November 7, 2002}

\abstract{(The image quality has been reduced to submit this paper to
astro-ph. To get a full resolution version, please visit {\tt 
http://www.astro.uu.nl/\~{}sluys/m1-67/} .)

A new point of view on the dynamics of the circumstellar nebula
M1-67 around the run-away Wolf-Rayet (WR) star WR~124 is presented.  
We simulated the outbursts of nebulae
with different morphologies, to compare the results to the
observed dynamical spectra of M1-67. We found that it has been
interacting with the surrounding ISM and has formed a bow shock
due to its high velocity of about $180~\mbox{km s}^{-1}$ relative to the
local ISM. The star is about 1.3 parsec away from the front of this
bow shock. The outbursts that are responsible for the nebula are
assumed to be discrete outbursts that occurred inside this
bow shock. The ejecta collide with this bow shock shortly after the
outburst.  After the collision, they are
dragged away by the pressure of the ISM, along the surface of the
bow shock. The bow shock is oriented in such way that we are looking from
the rear into this paraboloid, almost along the main axis.
Evidence for this is given firstly by the fact that the far
hemisphere is much brighter than the near hemisphere, secondly by
the fact that there is hardly any emission found with radial
velocities higher than the star's radial velocity, thirdly by the fact
that the star looks to be in the centre of the nebula, as seen
from Earth, and finally by the asymmetric overall velocity distribution of
the nebula, which indicates higher radial velocities in the
centre of the nebula, and lower velocities near the edges.
We find evidence for at least two discrete outbursts that occurred inside 
this bow shock.  For these outbursts, we find expansion velocities of
$v_\mathrm{exp} \approx 150~\mbox{km s}^{-1}$ and dynamical timescales of about 0.8 and $2 \times 10^4$~yr, 
which are typical values for LBV outbursts.  We therefore conclude
that M1-67 originates from several outbursts that occurred inside the bow shock
around WR~124, during an LBV phase that preceded the current WR phase of the star.


}

\maketitle

\section{Introduction}
\label{sec:01}

In this article, we describe our research on the dynamics of the
Wolf-Rayet ring nebula M1-67.  M1-67 is a bright nebula around the
Wolf-Rayet (WR) star WR~124. The star has a high heliocentric
velocity of almost 200~$\mbox{km s}^{-1}$ and is also known as 
\emph{\object{Merrill's star}} \citep{1938PASP...50..350M} and \object{209 BAC}. The star is
classified as a population~I WN8 star \citep{1964PASP...76..241B} and
is located in the constellation Sagittarius. Distance estimates
vary from about 4.5~kpc \citep{1979RMxAA...4..271P} to 6.5 kpc 
\citep{2000A&A...360..227N}. The
star has a terminal wind velocity of 710~$\mbox{km s}^{-1}$ and a mass
loss of $2.45 \times 10^{-5}~\mbox{M}_{\sun}~\mbox{yr}^{-1}$.  Its mass is
estimated to be about $20~\mbox{M}_{\sun}$ and its luminosity
$6 \times 10^5~\mbox{L}_{\sun}$ \citep{2000A&A...360..227N}.

The nebula M1-67 around WR~124 shows a clumpy structure, and most
of the gas is concentrated in knots and filaments 
\citep{1998A&A...335.1029S}.  An HST image of the nebula is displayed in
Fig.~\ref{fig:h3857_03}. The nebula was first classified as an
\ion{H}{ii} region. After the discovery that the nebula has about
the same radial velocity as WR~124, it was suggested that the
nebula might be a planetary nebula \citep{1946PASP...58..305M}.
However, the presence of a WR star and the N-enhancement and 
O-deficiency of the nebula suggest a WR ring nebula 
\citep{1998A&A...335.1029S}.  The distance estimates also point in the
direction of an ejected nebula, so that M1-67 is now generally
accepted as a Wolf-Rayet ring nebula.

Though WR ring nebulae are not necessarily ring-shaped, they
often exhibit a structure of arcs or rings.  This suggests that
the nebulae may be created by discrete outburst events.  It is
generally thought that WR ring nebulae originate from a Luminous
Blue Variable (LBV) stage of the central star, which is supposed
to precede the WR phase.

More than half of the LBVs have circumstellar nebulae 
\citep{1997lbv..conf..303N}. The different nebulae are very similar in
terms of physical properties. The expansion velocities are in the
order of 50 to $100~\mbox{km s}^{-1}$, their sizes about 1~parsec, and
the dynamical ages are in the order of $10^4$~yr. The
densities of the nebulae are generally found to be low (500 to 
1000~cm$^{-3}$) and the temperatures are in the range of 5000 to
10\,000~K \citep{2001ApJ...551..764L}.

It is still a point of debate whether the LBV outbursts occur
during a Red Supergiant (RSG) or a Blue Supergiant (BSG) phase.
The RSG scenario is proposed by 
\citet{1993ApJ...408L..85S,1996ApJ...468..842S}, 
who suggest that the ejection of mass occurs only
as a single event during a brief RSG phase. They explain the
enhanced abundances of heavy elements of the ejecta by convective
mixing in the RSG envelope 
\citep{1993ApJ...408L..85S,1996ApJ...468..842S}.

On the other hand, \citet{1994A&A...290..819L} suggest that
after the star has left the main sequence, it moves red-ward in
the Hertzsprung-Russel diagram (HR diagram) and the expanding
envelope becomes unstable, so that the star starts to develop
extreme mass loss.  This mass loss may be as high as 5 $\times$
10$^{-3}$ M$_{\sun}$ yr$^{-1}$ and is observed as the LBV ejecta.
They explain the chemically enriched ejecta by rotation induced
mixing. These eruptions therefore take place when the star is a
BSG and prevent it from becoming a RSG 
\citep{1994A&A...290..819L,2001ApJ...551..764L}.

The research of the chemical composition of LBV-ejecta by 
\citet{2001ApJ...551..764L} also indicates that the LBV eruptions
occur during a BSG phase. They suggest that the LBV outbursts are
induced by the rapid, near-critical rotation of the star.  In
their scenario, the star is also being prevented from becoming a
RSG by the mass loss. However, they point out that if a massive,
optically thick shell is being expelled from the star, it will
cool as it expands and the physical conditions will temporarily be
similar to that in the outer layers of a RSG, so that the
formation of dust can also happen in this case. This mechanism
explains the observed Humphreys-Davidson limit, that depicts the
lack of RSGs with luminosities higher than 6 $\times$ 10$^5$
L$_{\sun}$.

The goal of our research is to disentangle the geometry and
dynamics of the nebula M1-67. In order to do so, we create
different numerical models, of which the output is compared to
available observations.  Firstly, we present the observational
data we use for this study in Sect.~\ref{sec:02}. We
will then discuss our models for freely expanding outbursts in
Sect.~\ref{sec:03}. The reason why we let these models expand freely is that
the O-star that precedes a WR star blows a bubble of typically
30~pc in the ISM during its lifetime 
\citep{1999isw..book.....L}.  It will take an outburst of $100~\mbox{km s}^{-1}$
more than $10^5$~yr to cross this distance, so that it can
indeed be considered to expand without any disturbance.  We will
find that no satisfying fit can be made, so that the assumption
of a freely expanding outburst is wrong. We show that the cause for this is
that the star has a high velocity relative to the ISM.  This
causes a paraboloid-like bow shock instead of a more or less
spherical bubble. The star is about 1~pc away from the front of
this bow shock, so that an outburst of $100~\mbox{km s}^{-1}$ needs only
$10^4$~yr to cross it.  Once it has done so, part of the
outburst will collide with the bow shock, and will possibly be
dragged away along the bow shock surface. We discuss the
bow shock models for the case of a constant wind velocity in
Sect.~\ref{sec:04}. In Sect.~\ref{sec:05}
we compare the theoretical bow shock models for a constant stellar
wind with the observations and find a remarkable resemblance.
Sect.~\ref{sec:06} discusses the results of impacts of
outbursts on the bow shock surface.  In Sect.~\ref{sec:07}
we summarise the results and present the conclusions of this
study.

\section{Observations}
\label{sec:02}

For this research, we used the following three sets of
observational data.

\subsection{Long-slit spectra}
\label{sec:02.1}

\begin{figure*}
\sidecaption
\includegraphics[width=12cm]{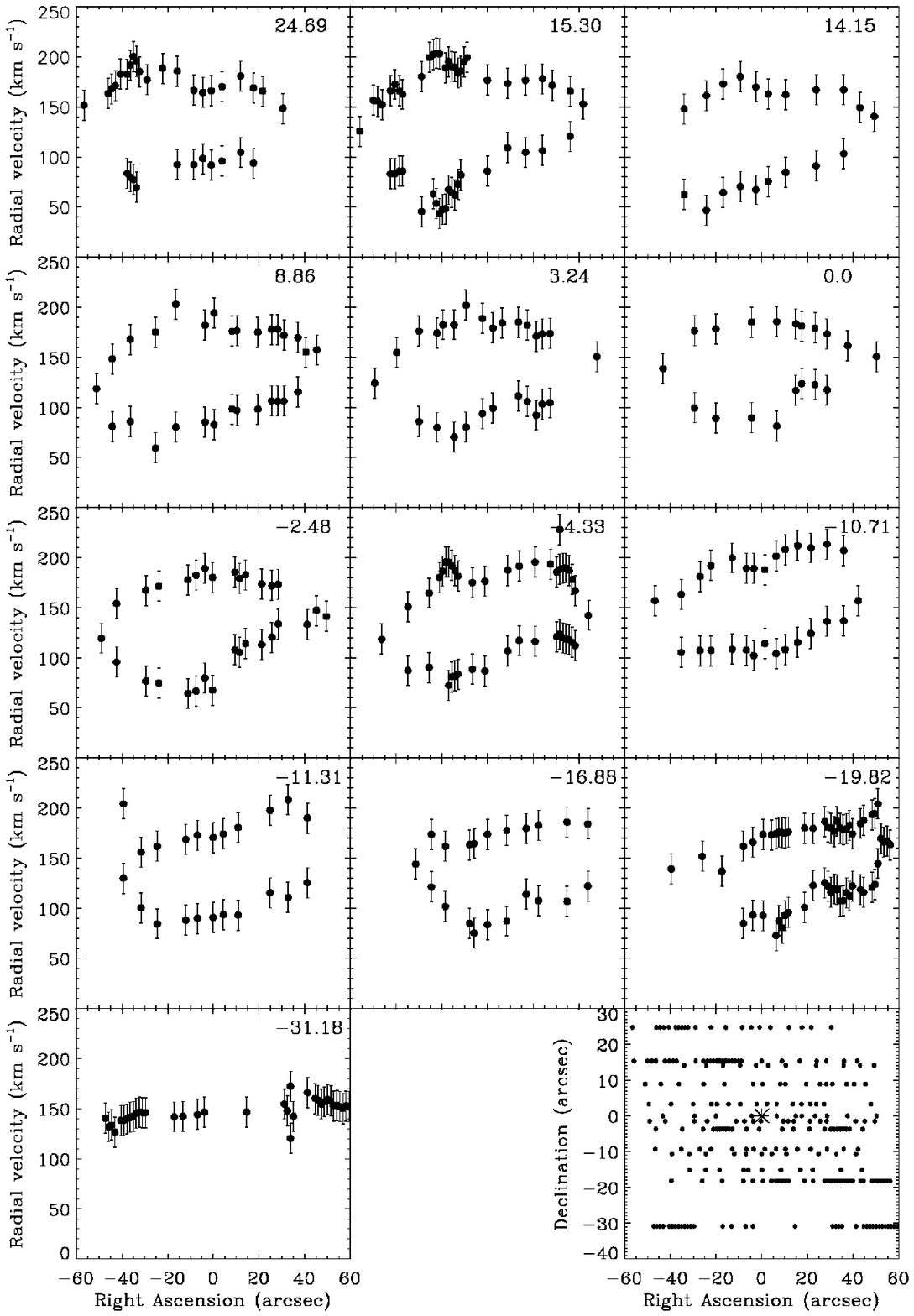}
 \caption{
 Position-velocity diagrams obtained from the observations by
 \citet{1998A&A...335.1029S}.  Each panel displays the result
 of one slit position, where the radial velocity is plotted
 against the position along the slit.  The lower right panel shows
 the coverage of data points in the sky, with the star in the
 origin, denoted by the asterisk.
        }
 \label{fig:h3857_01}
\end{figure*}

The first dataset we used, is velocity information from long slit 
spectra we obtained from
A. Nota, published in \citet{1998A&A...335.1029S}.  These data
consist of 13 long slit spectra, taken with the ESO Multi Mode
Instrument (EMMI) at the 3.5m NTT in La Silla.  Each slit is
positioned over the nebula in the east-west direction, at 
constant declination.  The declinations of the slits lie between
-30.82\arcsec and +24.69\arcsec relative to the star's declination.  In total, 413
good data points (right ascension, declination and radial
velocity) were derived from the spectra, which formed the input
for our study.  These data points have been plotted in
Fig.~\ref{fig:h3857_01}.

\subsection{Fabry-P\'erot images}
\label{sec:02.2}

\begin{figure*}
\sidecaption
\includegraphics[width=12cm]{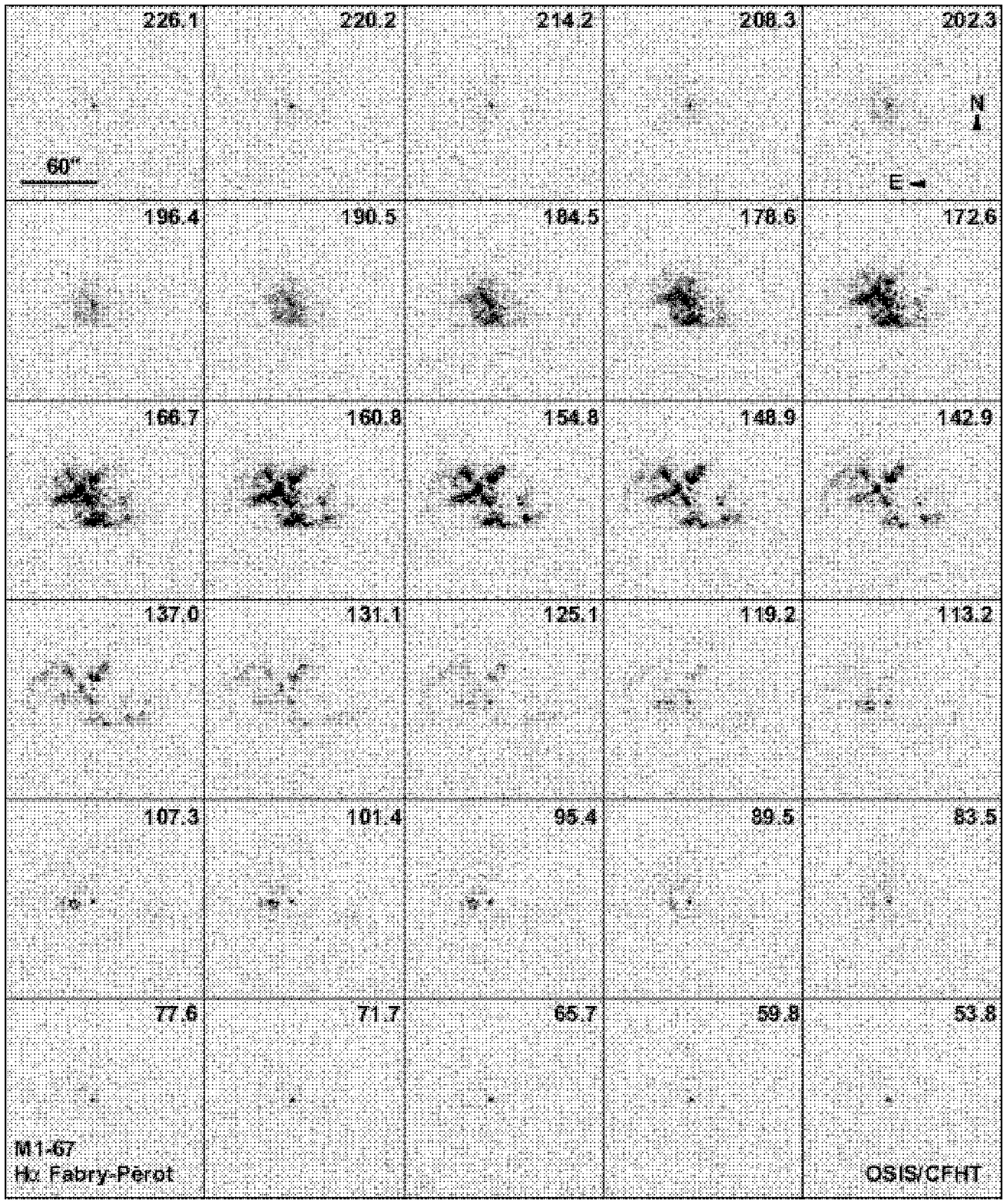}
 \caption{
 Fabry-P\'erot images obtained from \citet{1999npim.conf..453G}.  
 Each image shows a picture of the emission
 of M1-67 at a specific radial velocity, that is displayed in the
 upper right corner.  The star is in the centre of each image.  
        }
 \label{fig:h3857_02}
\end{figure*}

The second dataset are Fabry-P\'erot (FP) images obtained
from \citet{1999npim.conf..453G}. The FP images are shown in
Fig.~\ref{fig:h3857_02}. These 30 images were made in
August 1996, using CFHT-SIS, with the \'etalon of the Universit\'e
Laval, Qu\'ebec, Canada.  Each image was taken in H$\alpha$,
at a slightly different wavelength, so that it displays the
emission at a certain radial velocity. These observations give a
much more detailed view of the velocity distribution than the
long slit spectra.  Each image consists of 100 $\times$ 100
pixels.  From the darkest points, we used the coordinates ($x$,
$y$, $v$) in the same way as the data points from the long slit
spectra.  A few thousand points were used. Note that the southern
part of the nebula, more than approximately 20\arcsec south of the star, 
is missing (compare to Fig.~\ref{fig:h3857_03} and see Fig.~\ref{fig:h3857_08}(b))
due to deteriorating seeing during the observations 
\citep{1999npim.conf..453G}.

\subsection{HST image}
\label{sec:02.3}

For reference, we also used the Hubble Space Telescope image taken 
by \citet{1998ApJ...506L.127G}, for example to identify structures 
found in the Fabry-P\'erot images.

\begin{figure}
\resizebox{\hsize}{!}{\includegraphics{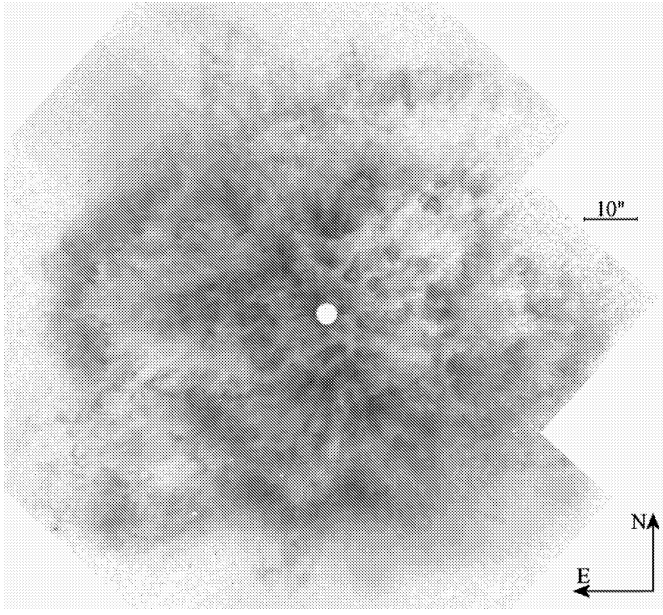}}
 \caption{
 Negative image of M1-67 made by the Hubble Space Telescope \citep{1998ApJ...506L.127G}.
    }
\label{fig:h3857_03}
\end{figure}

The image is a composite image of four WFPC2 images with a total 
exposure time of 10\,000 seconds, taken in March 1997, using the
narrow band F656N~H$\alpha$ filter.  In the same way, four images of 
the same field were taken through the broadband F675W~$R$ filter and 
combined to obtain a 'continuum' image close to H$\alpha$.  The continuum
image was flux-scaled and subtracted from the first composite image to obtain the 
deep continuum-subtracted H$\alpha$ image with the field stars removed, 
of which a negative version is shown in Fig.~\ref{fig:h3857_03}.

In this image different distinct arcs are visible. However,
although these arcs are clearly seen \emph{locally}, it is very
difficult to find a \emph{global} system of rings.  The fact that
the arcs are so clear, indicates that they are probably
formed during different, discrete outbursts, whereas the deficit
of a global pattern may tell us that the history of the nebula is
less straightforward than we might think a first.

 \section{Models of freely expanding outbursts}
 \label{sec:03}

 We have developed models to simulate ejected nebulae numerically
 and calculate spectra from them at different slit positions.
 The velocity information of the part of the nebula in
 the slit is then converted to a position-velocity plot
 (PV-plot), which can be compared to observational data.
 The purpose of these models is to give insight in the relation
 between features in PV-plots and the geometry of
 the nebulae that cause them.  Hence we have modelled the most
 common and most likely structures for ejected nebulae: spheres,
 ellipsoids and cones.

 \subsection{The geometry of the nebulae}
 \label{sec:03.1}
 We start with creating a geometry in the computer.  Our
 structures are hollow and made up of single dots that represent
 the surface of the nebula.  Solid or partially filled structures
 can be made by nesting different shells into each other.
   For each dot, that
 represents a small volume, the three rectangular
 coordinates with respect to the star are calculated.

 For \emph{spheres} we use:
 \begin{equation}
 \left. \begin{array}{lll}
 x &=& R_\mathrm{neb} \cos\psi \sin\theta \\
 y &=& R_\mathrm{neb} \sin\psi \sin\theta \\
 z &=& R_\mathrm{neb} \cos\theta
 \end{array} \right\}
 \label{eq:spheres}
 \end{equation}
 where $R_\mathrm{neb}$ is the radius of the nebula,
 $\psi$ ranges from $0$ to $2 \pi$ and $\theta$ from
 $-\frac{\pi}{2}$ to $+\frac{\pi}{2}$, in such way that the dots
 are evenly spread over the spherical surface.

 \emph{Ellipsoids} are made in the same way.  We only consider
 axially symmetric ellipsoids.  Their geometry is then defined by the oblateness
 $m \equiv R_\mathrm{pol}/R_\mathrm{eq}$,
 with $R_\mathrm{pol}$ and $R_\mathrm{eq}$ the polar and
 equatorial radius respectively.  The $x$ and $y$ coordinates are
 then similar to that of the sphere in Eq.~(\ref{eq:spheres}) and
 the $z$ coordinate is given by:
 \begin{equation}
 z = R_\mathrm{neb} \cos\theta \cdot m
 \label{eq:ellipsoids}
 \end{equation}

 \emph{Bipolar cones} are characterised by a semi-opening angle (SOA).  The
 nebulae with this geometry can be created using the equations:
\begin{equation}
 \left. \begin{array}{ll}
  x =& z \cos\psi \tan \mathrm{SOA} \\
  y =& z \sin\psi \tan \mathrm{SOA}
 \end{array} \right\}
\end{equation}
 where $z$ varies between $-R_\mathrm{neb}$ and
 $+R_\mathrm{neb}$ and $\psi$ between $0$ and $2\pi$.

\begin{figure}
\resizebox{\hsize}{!}{\includegraphics{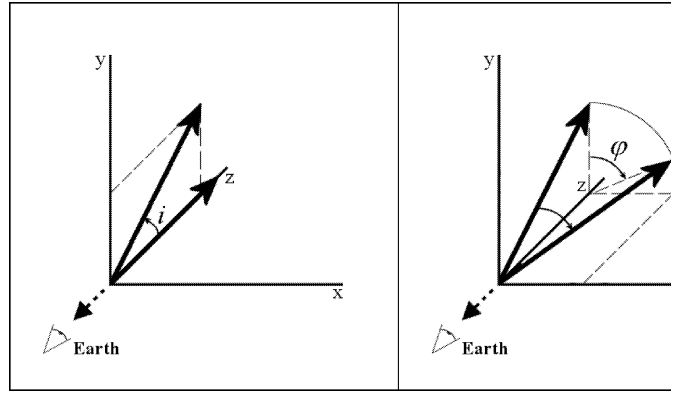}} \caption{
 Explanation of the coordinates and angles used in this section.
 The $x$ and $y$ axes represent the celestial coordinates right
 ascension and declination, the $z$ axis is the radial coordinate,
 with negative values toward the observer.  The thick arrow
 represents the symmetry axis of the structure.  \emph{Left
 panel:} the inclination is changed by rotation around the $x$ axis
 about the angle $i$.  \emph{Right panel:} the inclined structure is
 rotated around the $z$ axis to change the position angle $\varphi$.
 }
 \label{fig:h3857_04}
\end{figure}

 The ellipsoids and cones can be made
 either elongated to simulate jets, or flat to create
 disks.  All structures have axial symmetry and are created with
 the line of sight as their symmetry axis ($z$ axis).  They are then rotated
 around a line perpendicular to the line of sight ($x$ axis) to change the
 inclination $i$ and subsequently rotated around the line of sight
 (which is then no longer the symmetry axis) to change the
 position angle $\varphi$.  This is illustrated in
 Fig.~\ref{fig:h3857_04}.

 \subsection{The velocity models}
 \label{sec:03.2}

 After choosing the orientation of a structure, we overlay slits
 of finite width and transform the radial coordinates of the
 underlying points to radial velocities.  We assume the
 velocity-distance relation for the nebula to be
 \begin{eqnarray}
 v_\mathrm{r} & = & v_0\left(\frac{r}{R_\mathrm{neb}}\right)^{\alpha} \label{eq:velocity} \\
 v_\mathrm{z} & = & v_\mathrm{r}\left(\frac{z}{r}\right) = v_0\left(\frac{r}{R_\mathrm{neb}}\right)^{\alpha}
 \left(\frac{z}{r}\right)
\label{eq:vlos}
\end{eqnarray}
 Here, $R_\mathrm{neb}$ is the radius of the nebula, $v_0$ is
 the expansion velocity, $r = \left(x^2 + y^2 + z^2\right)^{1/2}$
 is the distance to the star and $z$ is the radial
 distance to the star, measured along the line of sight.  For
 $\alpha$ we can choose a suitable radius-velocity dependence:
 $\alpha < 0$ for braking after the outburst,
 $\alpha = 0$ for a continuous outflow with constant velocity,
 $\alpha = 1$ for a single, short burst with no further
 interaction and
 $\alpha > 1$ for acceleration of the material after the outburst. 
 For our models of freely expanding outbursts, a thin nebular shell with
 $\alpha = 1$ is used. Since all
 our structures are made up of single dots, the transformations can be
 carried out point by point.

 For each slit, our models produce one PV-plot, in which the radial velocity
 information under the slit is plotted against the position along
 the slit.  We took all slits parallel to the $x$ axis, so that they have
 a constant $y$ coordinate.  The PV-plots thus show the
 radial velocity $v_\mathrm{z}$ plotted against the position $x$.
 All plots are scaled to arbitrary
 units in position and velocity, because we are only interested in the shape of a structure.  In
 actual observations, the structures may therefore appear stretched
 out in either direction (position or velocity).
 The result of this whole exercise is a 'reference guide' that we used to
 determine the origin of certain structures in observed
 long slit spectra.\footnote{The simulated long-slit spectra for various geometries
 and velocities are available from the first author via e-mail.}

 \subsection{Freely expanding outbursts for WR~124}
 \label{sec:03.3}

 We tried to fit the different geometries described
 above and combinations of them to the long slit spectra data of M1-67,
 shown in Fig.~\ref{fig:h3857_01}.  The clumpy and chaotic structure
 of the nebula made it hard to fit.  We then also tried to fit
 parts of the nebula.  However, we never found a
 convincing solution.  We therefore questioned our assumption that
 the outburst around WR~124 is expanding freely and concluded that
 this may not be the case.  The Fabry-P\'erot data
 of Fig.~\ref{fig:h3857_02}, which we obtained afterwards, supported this conclusion.
 The following features in the velocity data of M1-67 particularly
 convinced us:
 \begin{itemize}
 \item There is hardly any nebular emission seen at radial velocities
 higher than +208 km s$^{-1}$, though the star has a radial velocity
 of about +199 km s$^{-1}$.  In case of a continuous outflow by a wind or discrete 
 LBV-like outbursts, one would expect to find nebular emission with radial velocities
 of tens, or even hundreds of kilometres per second higher than that of the star.
 \item Emission is seen at radial velocities from about +78 to
 $+208~\mbox{km s}^{-1}$, of which the average is $+143~\mbox{km s}^{-1}$.
 However, the bulk of this emission is found at
 velocities higher than about $+140~\mbox{km s}^{-1}$ (See
 Fig.~\ref{fig:h3857_02}).  This means that the
 high-velocity part of the nebula shows much more emission than the
 low-velocity part.  This was already mentioned by \citet{1982A&A...116...54S}.
 \item There is an asymmetry in the emission distribution. 
 When looking at the different panels in
 Fig.~\ref{fig:h3857_02}, starting at the highest velocity and
 skipping down, it is clearly seen that the emission appears
 close to the star and becomes increasingly broader.  When
 skipping from the lowest velocity panel upward, it is seen that
 (except for some small patches at lower velocities) the emission
 'starts' at about 120--$130~\mbox{km s}^{-1}$ as a very broad pattern.
 The far hemisphere is thus narrower than the near hemisphere and
 the bright emission region between +140 and $+208~\mbox{km s}^{-1}$
 looks like a triangle that is pointing away from us.
 \end{itemize}

 The explanation that can be found in literature is that the
 nebula as a whole is braked by the ISM, so that the star is displaced
 from the centre of expansion toward the leading edge
 \citep{1981ApJ...249..586C,1982A&A...116...54S}.  This means that the far
 and near hemisphere are braked equally, which is in contradiction
 with the asymmetry in the emission distribution.
 Instead, we think that the reason that the model of a freely
 expanding nebula does not hold for M1-67, is that there is indeed
 a strong interaction between the stellar wind and the ISM and that this
 interaction produces a bow shock around the stellar surface.  As
 we will point out in Sect.~\ref{sec:06}, the bow shock model is also
 qualitatively able to explain the chaotic structure of the
 nebula, which is much more difficult to do with a freely
 expanding nebula.

\section{The bow shock models}
\label{sec:04}

The failure of fitting a freely expanding shell to the velocity
data of M1-67, the high radial velocity of WR~124 and the global
picture from the more detailed Fabry-P\'erot data show that our
assumption of the freely expanding shell is incorrect. Instead,
the high velocity of the star is responsible for the formation of
a paraboloid-shaped bow shock. The possibility of a bow shock was 
already mentioned by \citet{1999npim.conf..453G}. The bow shock 
model allows the star to be much closer to the ISM, which makes it 
easier for outbursts to collide with it. We will discuss the models for such a
bow shock in this section. We start with a two-dimensional model
and convert it to a rotationally symmetric three-dimensional model, which can be
tilted to any wanted orientation, before obtaining the radial velocity
information.

\subsection{2D models for continuous mass loss}
\label{sec:04.1}

The models we use here are published by \citet{1996ApJ...469..729C}.  
They derive analytical expressions to describe
wind-wind interactions in general. They also discuss the special
case where one wind is plane-parallel and the other spherical, as
is the case when a star with spherical wind is ploughing through a
homogeneous ISM.  In our models we assume the interstellar medium to be homogeneous.  
We also assume that the stellar wind is constant over a long period ($\sim 10^4$~yr)
compared to the timescale of LBV-outbursts (usually on the order of 
$10^2$~yr), except during these outbursts.
The wind-ISM interaction is responsible for the geometry and
dynamics of the bow shock, while all emission comes from the
material that is ejected during the outbursts and its collision
into the bow shock surface.  This implies the assumption that
the momentum that is carried along with the outbursts is
lower than the momentum in the wind-ISM interaction.

\begin{figure}
\resizebox{\hsize}{!}{\includegraphics{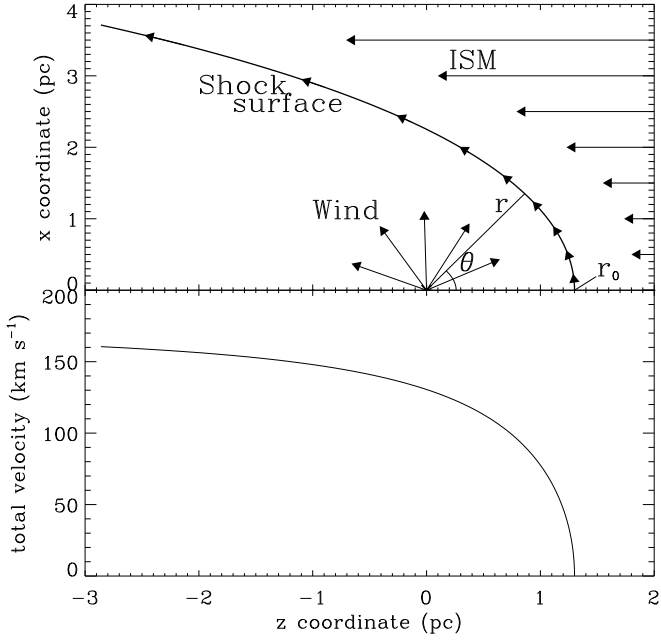}} \caption{
 Graphical representation of the bow shock model by 
 \citet{1996ApJ...469..729C}, with the physical parameters for WR~124.
 \emph{Upper panel (a):} geometry ($z$ versus $x$ coordinate), 
 \emph{lower panel (b):} velocity along the shock surface ($v$).  In both panels the
 horizontal axis represents the symmetry ($z$) axis of the shock in parsecs.
 }
 \label{fig:h3857_05}
\end{figure}

The distance between the star and the front of the bow shock is
called the stagnation point distance and can simply be derived by
momentum equilibrium:
\begin{equation}
r_0 = \left(\frac{\dot{M} v_\infty}{4\pi \rho_\mathrm{ism}
v_\mathrm{ism}^2}\right)^{1/2}
\label{eq:shockr0}
\end{equation}
where $v_\mathrm{ism}$ is the velocity of the ISM relative to the
star (we will observe the situation from the rest frame of the
star for simplicity). The geometry of the bow shock is expressed
in terms of $r$, the distance between the star and a point on the
bow shock, as a function of the angle $\theta$ that is defined in
Fig.~\ref{fig:h3857_05}(a):
\begin{equation}
r = r_0 \csc \theta \sqrt{3(1 - \theta \cot \theta)}
\label{eq:shockr}
\end{equation}
\citep{1996ApJ...469..729C}. The velocity of the material
along the surface of the shock is split in a component parallel
to the symmetry axis ($z$ axis, $v_\mathrm{z}$) and a component
perpendicular to it ($v_\mathrm{xy}$). They are given by:
\begin{eqnarray}
v_\mathrm{z} &=& \frac { \dot{M} v_\infty \sin^2\theta -
4\pi\rho_\mathrm{ism} v_\mathrm{ism}^2 r^2 \sin^2\theta}
{2\dot{M}(1-\cos\theta) + 4\pi\rho_\mathrm{ism} v_\mathrm{ism} r^2
\sin^2\theta}
\label{eq:shockvz} \\
v_\mathrm{xy} &=& \frac { \dot{M} v_\infty \left(\theta -
\sin\theta \cos\theta\right)} {2\dot{M}(1-\cos\theta) +
4\pi\rho_\mathrm{ism} v_\mathrm{ism} r^2 \sin^2\theta}
\label{eq:shockvr} \\
v &=& \sqrt{v_\mathrm{z}^2 + v_\mathrm{xy}^2} \label{eq:shockv}
\end{eqnarray}
\citep{1996ApJ...469..729C}, where $v$ is the total velocity
along the bow shock surface.
In Fig.~\ref{fig:h3857_05} the geometry and the total velocity
along the surface are displayed.

\subsection{3D models for continuous mass loss}
\label{sec:04.2}

The model of \citet{1996ApJ...469..729C} is two-dimensional
and can easily be converted into a 3D-model by
\begin{equation}
\left.\begin{array}{lll}
x &=& r \sin\theta \cos\psi \\
y &=& r \sin\theta \sin\psi \\
z &=& r \cos\theta \end{array} \right\}
\label{eq:3d}
\end{equation}
where $\theta$ is defined in Fig.~\ref{fig:h3857_05}(a) 
and $\psi$ takes evenly spread values
between 0 and $2\pi$ around the symmetry axis. Thus, the 3D
surface is obtained by rotating 
Fig.~\ref{fig:h3857_05}(a) around the horizontal axis. The velocity
components are given by
\begin{equation}
\left. \begin{array}{lll}
v_\mathrm{x} &=& v_\mathrm{xy} \cos\psi \\
v_\mathrm{y} &=& v_\mathrm{xy} \sin\psi
\end{array}\right\}
\label{eq:shockvxy}
\end{equation}
and by Eq.~(\ref{eq:shockvz}).

The three-dimensional bow shock model is now complete and oriented
in such a way that the observer is looking into the hollow shock from the
rear.  We can now rotate the shock to any orientation we like,
about the angles $i$ and $\varphi$ as indicated in
Fig.~\ref{fig:h3857_04}. The velocity components are rotated the
same way, and we can use the celestial coordinates $x$ and $y$
and the radial velocity $v_\mathrm{z}$ to compare the predicted radial
velocity maps to the observations.

\begin{figure*}
\resizebox{\hsize}{!}{\includegraphics{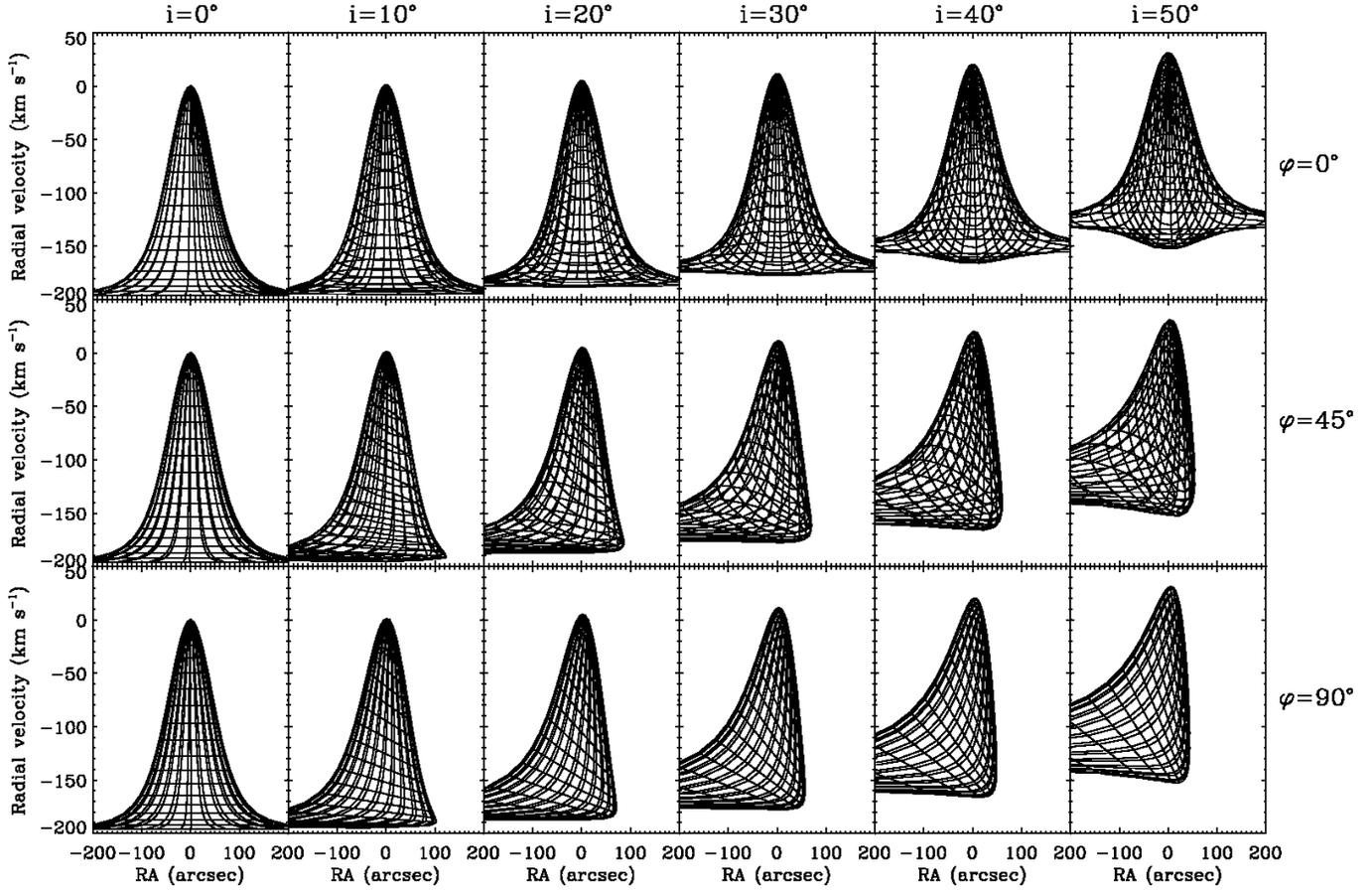}} \caption{
 Model output for the 3D bow shock models.  For all models
 $v_\mathrm{ism} = 200$ km s$^{-1}$ and the star is placed at a
 Right Ascension of $0\arcsec$ and a radial velocity of 0 km s$^{-1}$.
 The different rows have position angles ($\varphi$) of
 0, 45 and $90\degr$, the columns have inclinations ($i$) of 0, 10, 20, 30,
 40, and $50\degr$.  For each panel, the horizontal axis shows the
 Right Ascension (RA) in arcseconds, the vertical axis displays the
 radial velocity in km~s$^{-1}$.  See the text in Sect.~\ref{sec:04.2} for
 more explanation and interpretation.
 }
 \label{fig:h3857_06}
\end{figure*}

Fig.~\ref{fig:h3857_06} displays the Right Ascension against the
radial velocity $v_\mathrm{z}$ of a number of bow shocks with
different orientations.  From these plots, two effects of the
orientation of the bow shock can easily be seen.  Firstly, when
looking at the different columns from the left to the right, the
inclination of the bow shock increases.  The most important result
of this is that the maximum radial velocity observed shifts from
$0~\mbox{km s}^{-1}$ (which is the radial velocity of the star) for $i =
0\degr$ to more than $30~\mbox{km s}^{-1}$ for $i = 50\degr$. This
means that the velocity difference between the maximum radial
velocity where nebular emission is seen and the radial velocity of the
star puts an upper limit to the inclination.  As we have seen in
Sect.~\ref{sec:03.3}, in the case of M1-67 this
difference is approximately $10~\mbox{km s}^{-1}$, which gives a
strong indication for a small inclination of the main axis of the bow shock.  Secondly, when one
looks through the different rows from the top to the bottom, the
position angle changes.  The result is that the axial symmetry
around $\mbox{RA} = 0$ disappears.  We will later see that for the case
of M1-67, only about the upper $100~\mbox{km s}^{-1}$ in radial 
velocity are observed, so that
this effect manifests itself as a steep drop in radial velocity at 
one side of the maximum and a slower drop at the other side.  The side
of the steep drop indicates the direction where the top of the
bow shock points at.  For example, at the lower right plot of
Fig.~\ref{fig:h3857_06}, the steep side is at the right, which
means that the top of the bow shock is pointing to this side, so that the
star is moving, with respect to its local ISM, toward higher
Right Ascension.

\subsection{Orientation of the bow shock}
\label{sec:04.3}
\begin{figure}
\resizebox{\hsize}{!}{\includegraphics{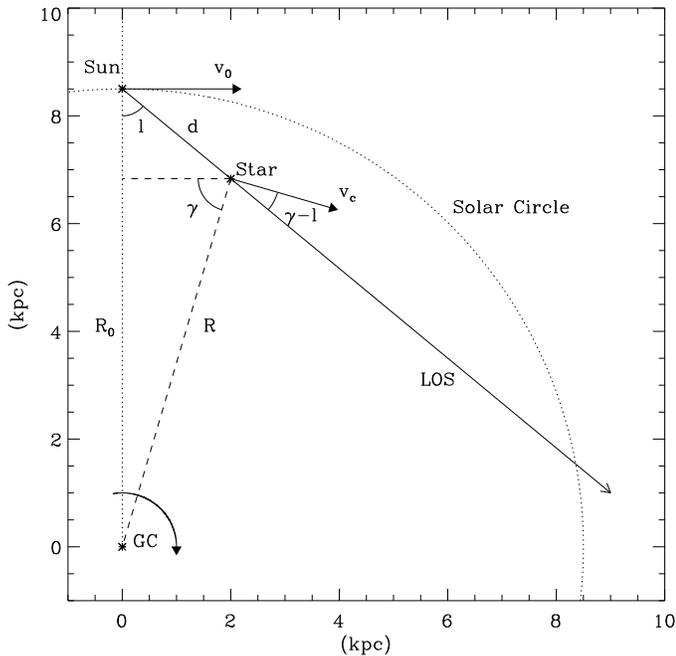}}
\caption
{
 The Milky Way as seen from the NGP, with the definitions of
 the quantities used in this section.
 }
 \label{fig:h3857_07}
\end{figure}

It is possible to roughly say something about the orientation of a
bow shock of a given star with respect to the line of sight (LOS).
This way, we can limit the total number of possible orientations of 
the bow shock in M1-67 appreciably.  In order to do so, we need 
to know the velocities of both the star and the ISM surrounding 
the star.  The orientation of the main axis of the bow shock is 
simply the orientation of the vector that is the difference of 
the spatial velocity of the star and that of the local ISM.  
The spatial velocity of the star can be derived from
the measured radial velocity and proper motion.  For the ISM
we will assume that it follows the laws of galactic rotation and
derive its velocity relative to the sun.  After subtracting the
ISM velocity from that of the star, we can calculate the total
velocity difference and the inclination and position angle of the
main axis of the bow shock. We will derive these quantities here
for the general case of a star with galactic longitude and
latitude $l$ and $b$ respectively and distance $d$ from the sun,
and then apply this to M1-67 in Sect.~\ref{sec:05}.

In the observations of the radial velocity and proper motion of
the star the motion of the sun is already taken into account, so
that we indeed need to correct for the motion of the sun when 
calculating the velocity of the ISM.  First, we will correct
for the motion of the local standard of rest (LSR), that moves
with a velocity of $v_0 = 220~\mbox{km s}^{-1}$ around the galactic
centre (GC) and subsequently the peculiar motion of the sun with 
respect to the LSR will be taken into account.  This peculiar 
velocity is calculated from Hipparcos observations for different 
classes of stars in the galactic plane by \citet{2000A&A...354..522M}. 
We will use the averages of the values given in that article:
\begin{equation}
\left. \begin{array}{lll}
u_{\sun} &=& 10.0 \pm 1.3 ~\mbox{km s}^{-1} \\
v_{\sun} &=& 13.9 \pm 3.7 ~\mbox{km s}^{-1} \\
w_{\sun} &=& 7.4 \pm 2.6 ~\mbox{km s}^{-1}
\end{array}\right\}
\end{equation}
for the direction toward the GC, the direction of galactic
rotation and toward the north galactic pole (NGP) respectively.

When looking along the line of sight (LOS) defined by the
galactic coordinates $l$ and $b$, we can calculate for a point at
any distance from the sun $d$, the distance from that point to the
GC, which we will call $R$ (see Fig.~\ref{fig:h3857_07}):
\begin{equation}
R = \left(d^2 \sin ^2 l + \left(R_0 - d \cos l\right)^2 \right)^{1/2}
\end{equation}
where $R_0$ is the standard value for the distance from the sun to
the GC: $R_0 = 8.5$ kpc. We can then calculate the angle
$\gamma$, defined in Fig.~\ref{fig:h3857_07}, by
\begin{equation}
\gamma = \arctan \left(\frac{R_0 - d \cos l}{d \sin l}\right)
\end{equation}
The circular velocity for a distance $R$ with 3~kpc $<$ $R$ $<$
$R_0$ is given by \citet{1988gera.book..295B}:
\begin{equation}
v_\mathrm{c} = v_0 \left( 1.0074
\left(\frac{R}{R_0}\right)^{0.0382} + 0.00698\right)
\end{equation}
The components of the ISM velocity in the radial direction and
the direction of galactic longitude and latitude, are then given
by:
\begin{equation}
\left. \begin{array}{lll}
v_{\mathrm{ism,r}} &=& v_\mathrm{c} \cos(\gamma - l) \cos b \\
v_{\mathrm{ism,l}} &=& v_\mathrm{c} \sin(\gamma - l) \cos b \\
v_{\mathrm{ism,b}} &=& v_\mathrm{c} \sin b
\end{array}\right\}
\label{eq:ismvelocity}
\end{equation}
In these expressions, we have neglected the velocity component of the
ISM perpendicular to the plane of the galaxy. 

The proper motion
of the star can be expressed in terms of galactic coordinates 
($\mu_l$ and $\mu_b$).  The proper motion gives rise to spatial 
velocity components that are dependent of the distance. We thus have:
\begin{equation}
\left.\begin{array}{lll}
v_\mathrm{*,r} && \\
v_\mathrm{*,l} &=& \mu_l \cdot d \\
v_\mathrm{*,b} &=& \mu_b \cdot d
\end{array}\right\}
\end{equation}

We can now calculate the velocity difference between the star and
the local ISM in the three
components, where we also correct for the motion of the sun:
\begin{equation}
\left. \begin{array}{lll}
v_\mathrm{r} &=& v_\mathrm{*,r} - \left( v_{\mathrm{ism,r}} - v_0 \sin l - u_{\sun} \cos l - v_{\sun} \sin l \right)\\
v_\mathrm{l} &=& v_\mathrm{*,l} - \left( v_{\mathrm{ism,l}} - v_0 \cos l + u_{\sun} \sin l - v_{\sun} \cos l \right)\\
v_\mathrm{b} &=& v_\mathrm{*,b} - \left( v_{\mathrm{ism,b}} + w_{\sun} \right)
\end{array}\right\}
\label{eq:totalvelocity}
\end{equation}

The total velocity of the star with respect to the local ISM
is what we called $v_\mathrm{ism}$ in Sect.~\ref{sec:04.1}:
\begin{equation}
v_\mathrm{ism} = \left( v_\mathrm{r}^2 + v_\mathrm{l}^2 +
v_\mathrm{b}^2 \right)^{1/2} \label{eq:bsvelocity}
\end{equation}
Furthermore, we can also find an expression for the inclination
$i$ and position angle $\varphi$:
\begin{eqnarray}
i &=& \arccos\left(\frac{v_\mathrm{r}}{v_\mathrm{ism}}\right)
\label{eq:bsinclination} \\
\varphi &=& \arctan\left(\frac{v_\mathrm{l}}{v_\mathrm{b}}\right)
+ \epsilon \label{eq:bspositionangle}
\end{eqnarray}
where $\epsilon$ denotes the angle between the line of constant
declination and the line of constant galactic latitude at the
position of the star.

The uncertainties in the distance and in the proper motion of the 
star introduce a range of possible values for $v_\mathrm{ism}$, $i$ 
and $\varphi$, which is a constraint that we can use to reduce the
total number of possible orientations drastically.  This makes it 
easier to compare our models to the observations, as we will do in 
the next section.

\section{Comparison of bow shock models for continuous mass loss with observations}
\label{sec:05}

\begin{figure}
\resizebox{\hsize}{!}{\includegraphics{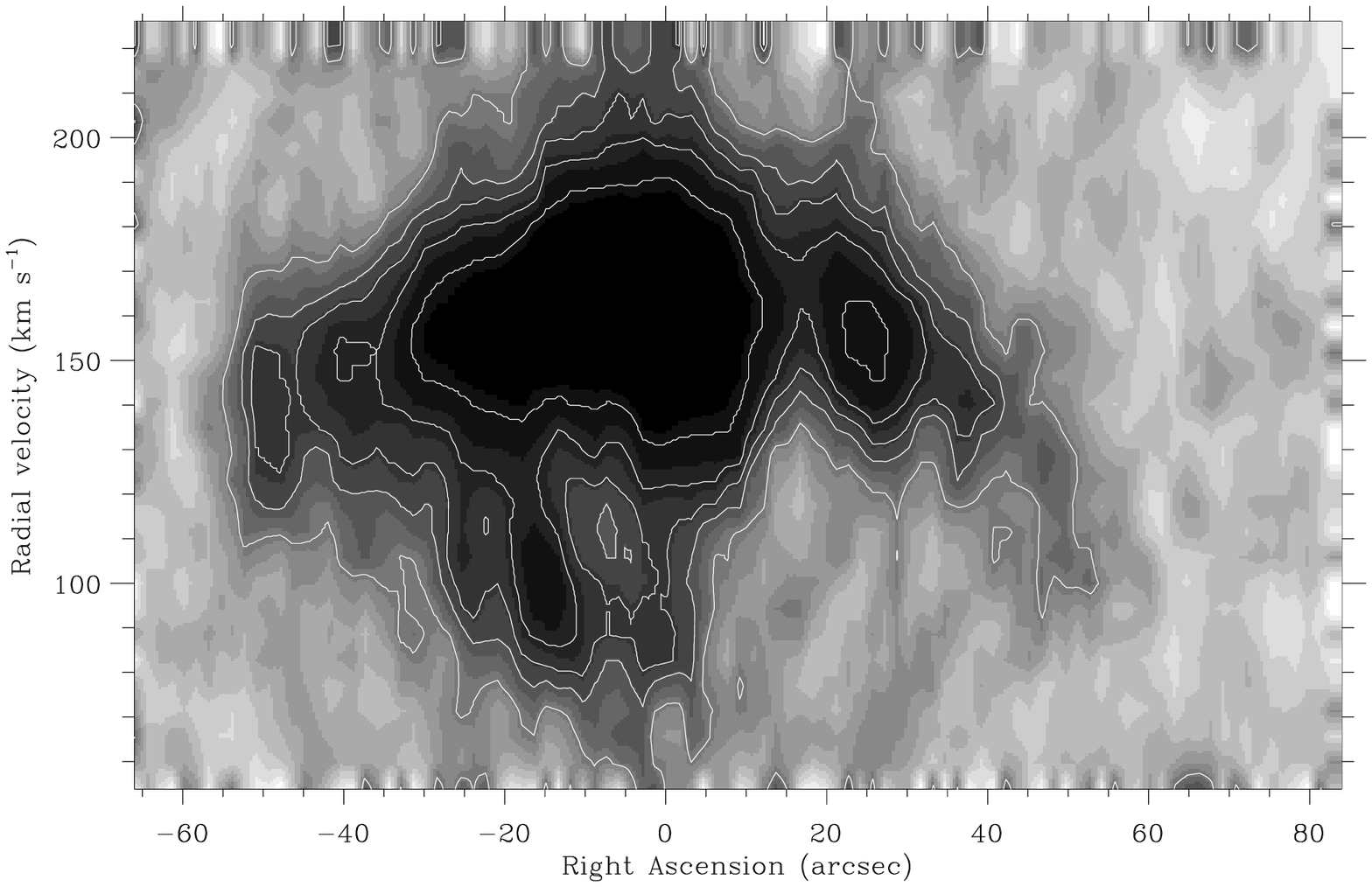}}
\resizebox{\hsize}{!}{\includegraphics{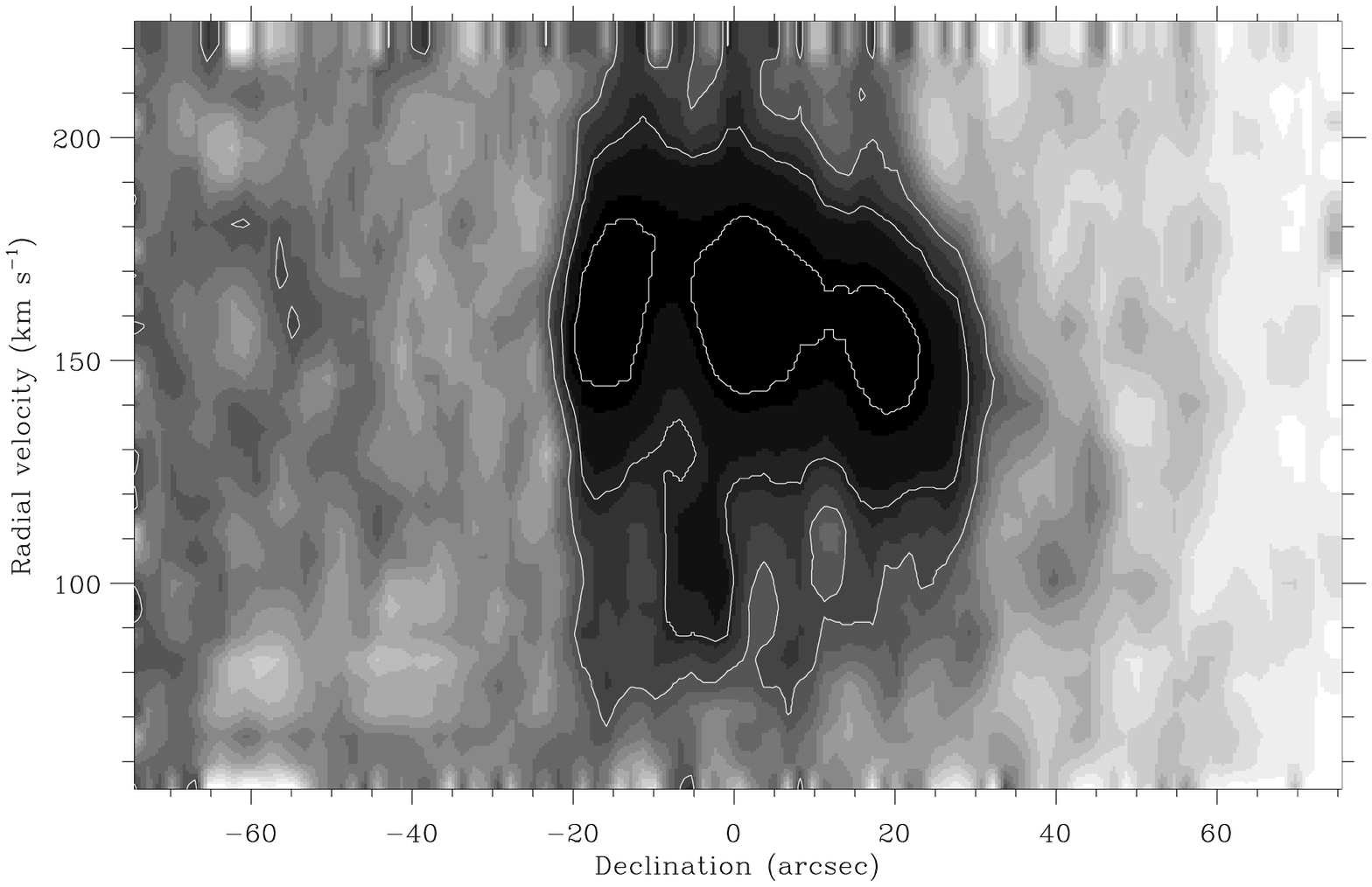}} \caption{
 Negative images of the emission distribution as inferred from the
 Fabry-P\'erot images in Fig.~\ref{fig:h3857_02}.  \emph{Upper
 panel (a):} radial velocity against Right Ascension, \emph{Lower
 Panel (b):} radial velocity against Declination.  The process
 of creating this image is described in the text.
 The disturbances in the edges of the image are artifacts caused
 by manipulating the image.  The sharp cutoff in the lower image
 at about -20\arcsec is caused by the seeing during the observations
 (see Sect.~\ref{sec:02.2}).
 }
 \label{fig:h3857_08}
\end{figure}

In this section we will use the bow shock model discussed in
Sect.~\ref{sec:04} to explain the distribution of emission as
observed in the Fabry-P\'erot images.  In order to do this
properly, we transformed the observational data in Fig.~\ref{fig:h3857_02} to
an image that displays the emission distribution in the Right
Ascension -- radial velocity plane.  This was done by stacking
the 30 different FP images on top of each other to create a 3D
body, with two spatial dimensions and one radial velocity
dimension. Then we summed all layers with constant declination to
get the image that is shown in Fig.~\ref{fig:h3857_08}(a). 
Fig.~\ref{fig:h3857_08}(b)
shows the same image, but here the Right Ascension is replaced by
the declination. These are the images we will compare our
bow shock models with different parameters to.

\begin{figure}
\resizebox{\hsize}{!}{\includegraphics{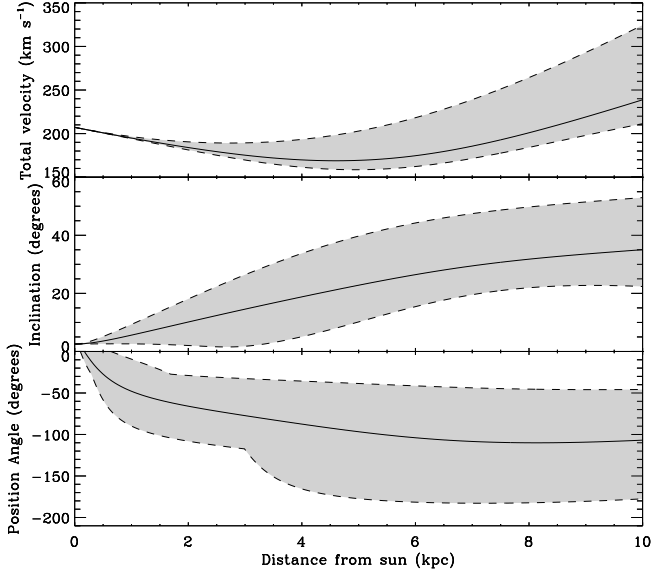}} \caption{
 Theoretical properties of the orientation of the bow shock around WR~124 as a
 function of the distance from the sun:
 \emph{Upper panel:} Total velocity of WR~124 relative to the ISM.
 \emph{Middle panel:} Inclination ($i$) of the main axis of the bow shock.
 \emph{Lower panel:} Position Angle ($\varphi$) of the main axis of the bow shock.
 The solid line gives the most probable value, the dashed lines display the
 uncertainties that are a result of the uncertainty in the proper motion of the star.}
 \label{fig:h3857_09}
\end{figure}

We apply the properties of WR~124 to the results that we
derived in Sect.~\ref{sec:04.3}. The star has galactic
coordinates $l = 50.2\degr$, $b = +3.31\degr$.  At this
position, the angle $\epsilon$ has the value of $62.7\degr$. The
proper motion of WR~124 was measured by Hipparcos and can be
converted to galactic coordinates:
\begin{equation}
\left. \begin{array}{lll}
\mu_l &=& -6.1 \pm 2.0 ~\mbox{mas yr}^{-1} \\
\mu_b &=& -3.0 \pm 2.0 ~\mbox{mas yr}^{-1}
\end{array}\right\}
\end{equation}

We can then use Eqs.~(\ref{eq:bsvelocity}), (\ref{eq:bsinclination})
and (\ref{eq:bspositionangle}) to calculate the velocity difference
between the star and the ISM ($v_\mathrm{ism}$), as well as the
inclination ($i$) and position angle ($\varphi$) of the main axis
of the bow shock (as defined in Fig.~\ref{fig:h3857_04}), as a
function of the distance ($d$) of WR~124 from the sun.  The results
are plotted in Fig.~\ref{fig:h3857_09}, where the most probable
value, the upper and lower limit of the proper motion of M1-67 are
depicted in the solid and dashed lines respectively. Because of the great
distance to the star, the uncertainties in the proper motion are
quite large, so that the resulting uncertainties in the velocity
components perpendicular to the LOS are also large and increase
with the distance.

Next, we apply the results from Sect.~\ref{sec:04.1}.  For
$d = 6.5$~kpc, Eq.~(\ref{eq:bsvelocity}) gives $v_\mathrm{ism}
\approx 174~\mbox{km s}^{-1}$ and we find from Eq.~(\ref{eq:shockr0})
that $r_0 = 1.3$~pc, where we used $\dot{M} = 2.45 \times
10^{-5}~\mbox{M}_{\sun}~\mbox{yr}^{-1}$, $v_\infty = 710~\mbox{km s}^{-1}$ 
\citep{2000A&A...360..227N} and 
$\rho_\mathrm{ism} = 1~m_\mathrm{H}~\mbox{cm}^{-3}$.

\begin{figure*}
\resizebox{\hsize}{!}{\includegraphics{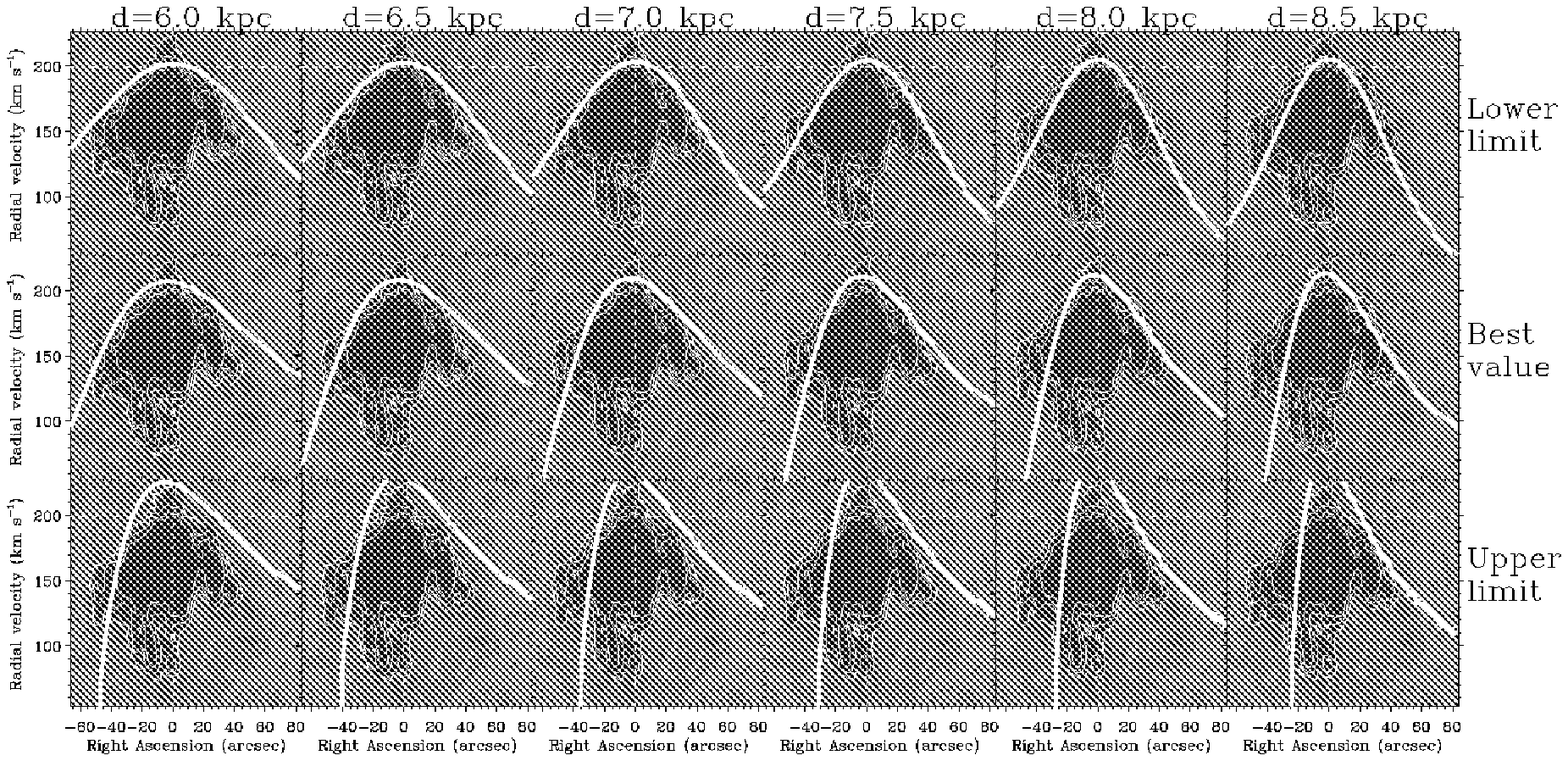}}
\resizebox{\hsize}{!}{\includegraphics{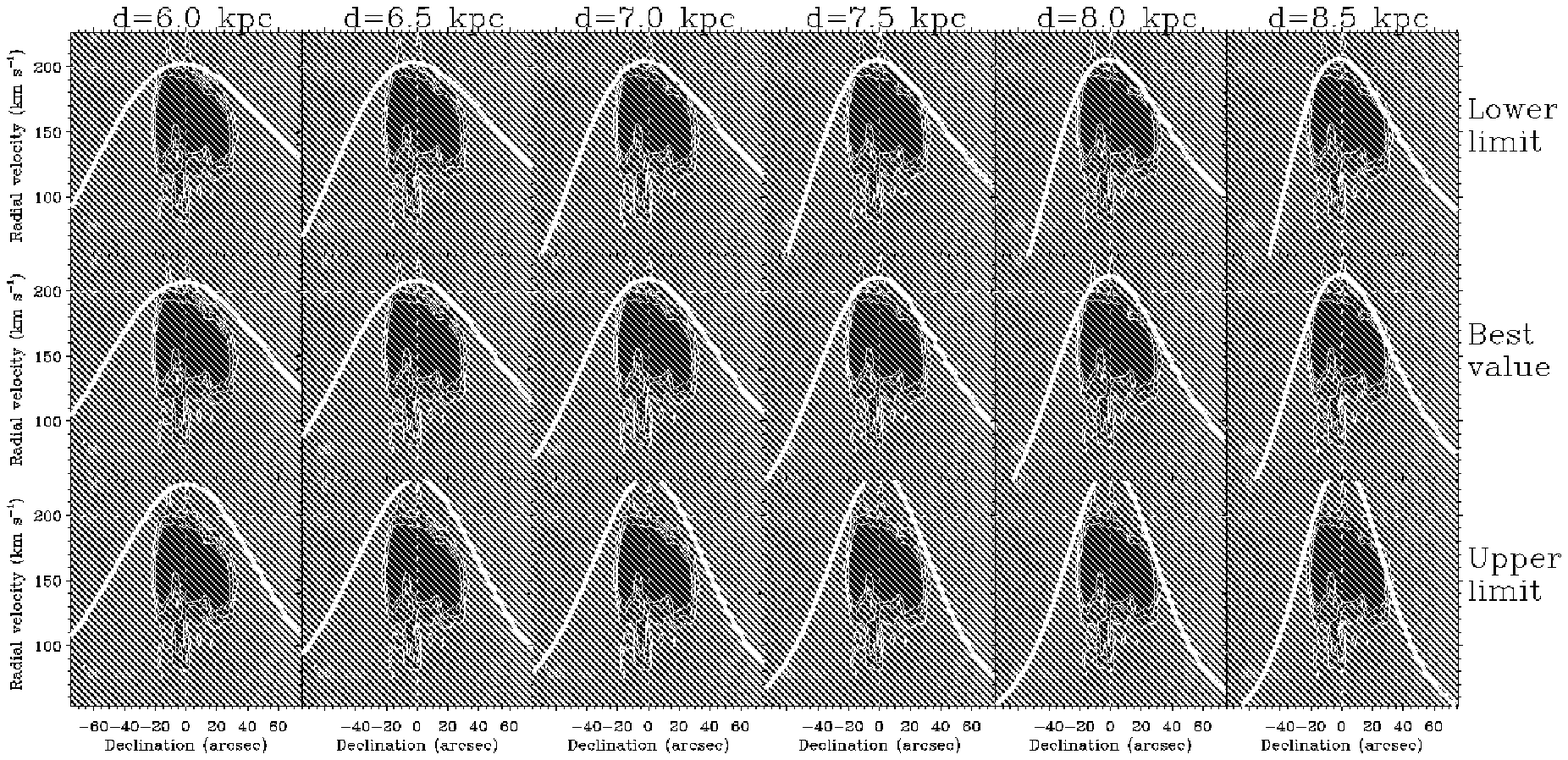}} \caption{
 Model output for the 3D bow shock models for WR~124.  
 \emph{Upper panel (a):} Right Ascension (\arcsec) versus radial 
 velocity (km~s$^{-1}$),
 \emph{lower panel (b):} Declination (\arcsec) versus radial 
 velocity (km~s$^{-1}$).
 The grey-scaled images in the background are smoothed 
 versions of Fig.~\ref{fig:h3857_08}.  The
 thin lines are contours of the data.  The model
 output is plotted as the thick curve, that encloses the actual
 bow shock in this projection and that would look like the output
 in Fig.~\ref{fig:h3857_06} if plotted completely.
 The different columns have values for the total velocity,
 inclination and the position angle for the distances of 6.0, 6.5,
 7.0, 7.5, 8.0 and 8.5 kpc. The rows are the three different
 values for these quantities for the lower limit, the best value
 and the upper limit for the proper motion respectively, so that each
 row displays one line of each graph in
 Fig.~\ref{fig:h3857_09}. The dashed lines in each plot are 
 at $0\arcsec$ and $199~\mbox{km s}^{-1}$,
 the position and radial velocity of WR~124.
  Note the lack of observational data at declinations lower than
 -20\arcsec (See Sect.~\ref{sec:02.2}).
 }
 \label{fig:h3857_10}
\end{figure*}

We can now build the 3D models as described in
Sect.~\ref{sec:04.2} and rotate them according to the
angles found in Fig.~\ref{fig:h3857_09}.  Because of the
large uncertainty in the proper motion, we calculate three
different models for each distance $d$, with the three different 
values found at each line in Fig.~\ref{fig:h3857_09} for 
$v_\mathrm{ism}$, $i$ and $\varphi$
and plot a contour of the model output over Fig.~\ref{fig:h3857_08}. The result
is displayed in Fig.~\ref{fig:h3857_10}, for distances
of 6.0, 6.5, 7.0, 7.5, 8.0, and 8.5~kpc.

The resemblance between the models and the data is remarkable. It
is clear that the best model can be found somewhere in the
first row and the third or fourth column of
Fig.~\ref{fig:h3857_10}.  From that, it can be inferred that the
best values for the parameters are $v_\mathrm{ism} \approx
180~\mbox{km s}^{-1}$, $i \approx 20\degr$ and $\varphi \approx
-185\degr$, so that the star is moving south, under a very small
inclination with respect to the line of sight. The resemblance
between the data and the model indicates that the values for these
three parameters fit well, not that this distance is better than
other values. The distance was only used to obtain likely values
for these parameters.  As can be seen from
Fig.~\ref{fig:h3857_09}, there are more distances that can
explain this combination of the three parameters. We conclude that the
bow shock model with the parameters given above is successful in
explaining the overall features of the two velocity versus position
($\alpha$ or $\delta$) plots of Fig.~\ref{fig:h3857_08} and 
Fig.~\ref{fig:h3857_10}.

There are of course also differences between the model output and
the observations. Part of these differences can be attributed to 
the fact that we assume a continuous mass loss from the star, which 
is in reality unlikely to be the case. In fact, in Sect.~\ref{sec:06}
we will argue that the mass loss occurred in outbursts.
Also, we assume the ISM
to be homogeneous.  Inhomogeneities in the ISM will cause a less smooth
bow shock surface, as indeed seen in the observations.  The largest
disagreement between model and observations is of course seen in 
the lower panel of Fig.~\ref{fig:h3857_10}.  This is due to the lack 
of observational data for declinations less than about -20\arcsec 
(see Sect.~\ref{sec:02.2} and Fig.~\ref{fig:h3857_08}).

\section{The effect of outbursts on the bow shock}
\label{sec:06}

So far, we have looked at a bow shock that is formed due to a
continuous wind.  However, as we have already noted in
Sect.~\ref{sec:01} and \ref{sec:02.3}, the distinct arcs
that can be seen in the HST image are likely to originate from
discrete outbursts.  Since these kinds of outbursts only take
place after the star has moved from the main sequence and since the O-star
that preceded WR~124 must already have had a strong wind, these
outbursts must have occurred \emph{inside} the bow shock.  In this
section we will describe what the effects of a short outburst on
the bow shock is.

In the first phase the outburst will expand freely and both the
geometrical and velocity structure are undisturbed,
apart from the extra radial velocity due to the velocity of the star.
Typical timescales for this phase are derived from
division of the distance $r_0$ of a few parsec by a typical LBV
outburst velocity of $100~\mbox{km s}^{-1}$, which gives a few times
$10^4$~yr.  After this time, the outburst will impact onto
the bow shock surface, which will affect the observed nebula in
different ways.  Firstly, the outburst will be braked or possibly
even be halted by the surface of the bow shock.  This means
that the radial velocity will change drastically and that the
nebula will brighten.  Even in the case of a spherical outburst,
the different parts of the outburst will reach the bow shock at
different times, so that these effects of braking and brightening
will propagate through the
nebula, starting at its top.  Secondly, the gas of the outburst
will be dragged along the surface of the bow shock and eventually
adopt its velocity.  This means again a difference in the
velocity pattern and a distortion in the (thus far possibly 
symmetric) geometry of the outburst.

Because the magnitude of the effect described above will depend
strongly on the relative momenta of the matter in both the
bow shock and the outburst, we divide the outbursts in three
simple categories:
\begin{enumerate}
\item Spherical outbursts where the momentum of the outburst is
smaller than that of the wind-ISM shock
\item Spherical outbursts where the momentum of the outburst is
larger than that of the wind-ISM shock
\item Non-spherical outbursts that may be a combination of the
two cases above: in some directions $p_\mathrm{outburst} <
p_\mathrm{bow shock}$, in others $p_\mathrm{outburst} >
p_\mathrm{bow shock}$. 
\end{enumerate}

In the first case, the geometry and dynamics of the bow shock will
not be changed too much by the impact of the outburst. Those
of the outburst, however, will change appreciably. The freely expanding
gas will impact onto the bow shock, brighten and be dragged along
the surface layer of the bow shock.  The outburst will adopt the
geometry and dynamics of the bow shock quickly, so that it is
eventually only recognisable as a bright, distorted ellipse in the
diffuse, weak, if at all visible, background of the gas that resides
in the bow shock.

In the second case, the steady bow shock is distorted by the
impact.  The outburst will be brightened and slowed down at the 
moment of impact, but will continue its motion for a while.  Eventually, the outburst
will be halted by the ongoing ram-pressure of the ISM, but this will
happen at a larger distance from the star than where it first hit 
the bow shock. This situation can be described in the same way 
as the situation before the outburst, only with a
stronger wind (with lower velocity but a much higher mass loss).
This will result in a larger stagnation point distance $r_0$, as
given by Eq.~(\ref{eq:shockr0}).  The outburst will change the
shape of the bow shock temporarily, because the different parts of
the outburst will impact at different times.  The distortion will
thus propagate through the bow shock, starting at its top.
Eventually, this results in a bow shock with the same geometry,
but a different size and different dynamics.  The size simply
scales with $r_0$, but the velocities in Eq.~(\ref{eq:shockvz}) and
Eq.~(\ref{eq:shockvr}) depend on the mass loss and terminal wind
velocity in a more complex way. After the outburst, the  
momentum of the quiet wind will drop back
to its old value, because the outburst itself lasts only very
short compared to the evolutionary timescale of the star in this phase.
Because of this, the old value of $r_0$ may be reinstalled 
after some time.

In the third case a non-spherical outburst occurs.  Many LBV
outbursts are known to be bipolar, the best known example
being $\eta$ Carinae.  Bipolarity of an LBV outburst
can result from binarity, or from fast rotation.  Since an interacting
spherical outburst as described above can at best result in an elliptical 
ring, the structures in this case are already more complex.  This
complexity will increase when the impact occurs. In
the case of a non-spherical outburst, it is likely that the outburst will not 
hit the bow shock at its top first, but somewhere on the side, where the
surface velocity of the shock is much higher.  Thus the resulting
drag will be stronger and will decrease symmetry even further.

\begin{figure}
\resizebox{\hsize}{!}{\includegraphics{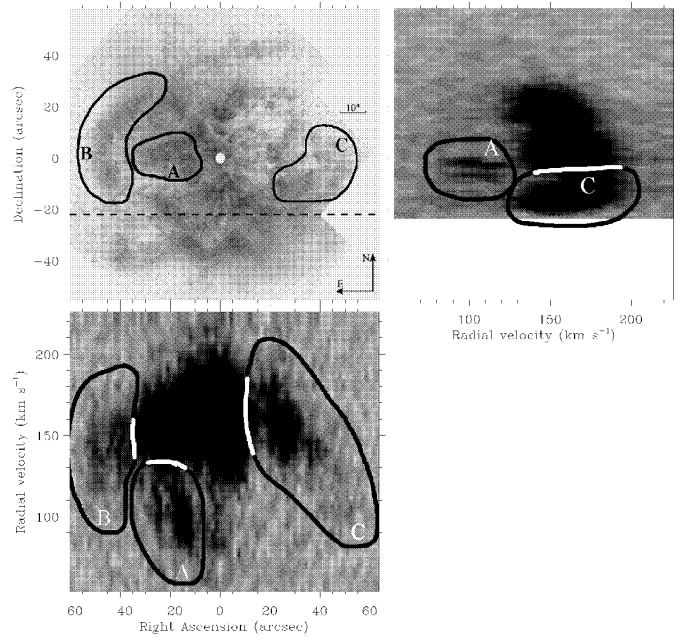}}
\caption{ The two data sets used for this research.
 \emph{Upper left panel (a):} HST image, \emph{upper right panel (b):} Fabry-P\'erot data 
 projected in the declination-radial velocity plane, \emph{lower left panel (c):} 
 Fabry-P\'erot data projected in the right ascension-radial velocity plane.  
 In the images, parts of the nebula that are discussed in the text are encircled 
 and labelled, for easier reference.  The reader can also easily compare the positions
 of each structure in the different images. Note that (roughly) the part of the upper 
 right image where no data is available is left blank.
 }
 \label{fig:h3857_11}
\end{figure}

In all cases, part of the outburst will move toward the rear end of the 
bow shock.  Because this end is open, it will never interact with the bow shock
surface and keep expanding 
freely.  This part of the outburst can therefore be used to derive a dynamical 
timescale directly.  In the velocity data of M1-67, we find a freely expanding 
structure that is labelled as \emph{A} in Fig.~\ref{fig:h3857_11}.  From comparison with 
Fig.~\ref{fig:h3857_10} it is clear that this structure is not located on the
bow shock surface.  We fitted the structure \emph{A} to an ellipse, assuming that 
WR~124 is at the centre of expansion. From the fit, we found $v_\mathrm{exp} \approx 
150 \pm 15~\mbox{km s}^{-1}$ and $r \approx 40\arcsec$ ($\propto~1.3$~pc at $d = 6.5$~kpc), which 
gives a dynamical timescale of $1250 \pm 125~\mbox{yr}~\mbox{kpc}^{-1}$ or about $8.2 
\pm 0.8 \times 10^3$~yr for for the estimated distance of 6.5~kpc for WR~124.  
This timescale is typical for LBV 
outbursts and therefore a strong hint that the nebula M1-67 may indeed be the 
result of LBV-like eruptions.  It is very unlikely that the structure \emph{A} is the
only result of the outburst that caused it, which we will refer to as outburst \emph{A}.  
The structure is clearly expelled toward us, and one would expect to find a 
counterpart of this structure that was caused by the same outburst, but directed 
away from us.  This counterpart should be clearly visible, as long as it hasn't reached
the bow shock surface yet, because it should have a much higher radial velocity than
the star (up to $+345~\mbox{km s}^{-1}$). Since we do not see this counterpart, it must have 
collided with the bow shock surface already. Given the expansion velocity derived above
($150~\mbox{km s}^{-1}$) and the stagnation point distance of $r_0 \approx 1.3$~pc 
(Sect.~\ref{sec:05}), it would take the part of outburst \emph{A} that is directed 
away from us about $8.5 \times 10^3$~yr to reach the bow shock surface.  Within the 
accuracy of the fit, this time agrees with the fitted dynamical timescale of $8.2 
\pm 0.8 \times 10^3$~yr.  This means that the counterpart 
of the freely expanding structure \emph{A} might have collided with the bow shock
just a short time ago. It is therefore very likely that the structure \emph{A} originates from the 
most recent outburst that occurred on WR~124.

The interactions between parts of an outburst and the steady bow shock can 
cause a very chaotically looking nebula, just as is the case for M1-67. In
particular, it will generate elliptical structures of which many can be seen
in Fig.~\ref{fig:h3857_03}.  
Another result of the impact of an outburst onto the bow shock, is that it is 
no longer possible to simply fit an undisturbed, freely expanding nebula to 
this part and derive its dynamical timescale.  However, we may still be able
to estimate the dynamical timescales of parts of the outburst that have been 
interacting.

\begin{figure}
\resizebox{\hsize}{!}{\includegraphics{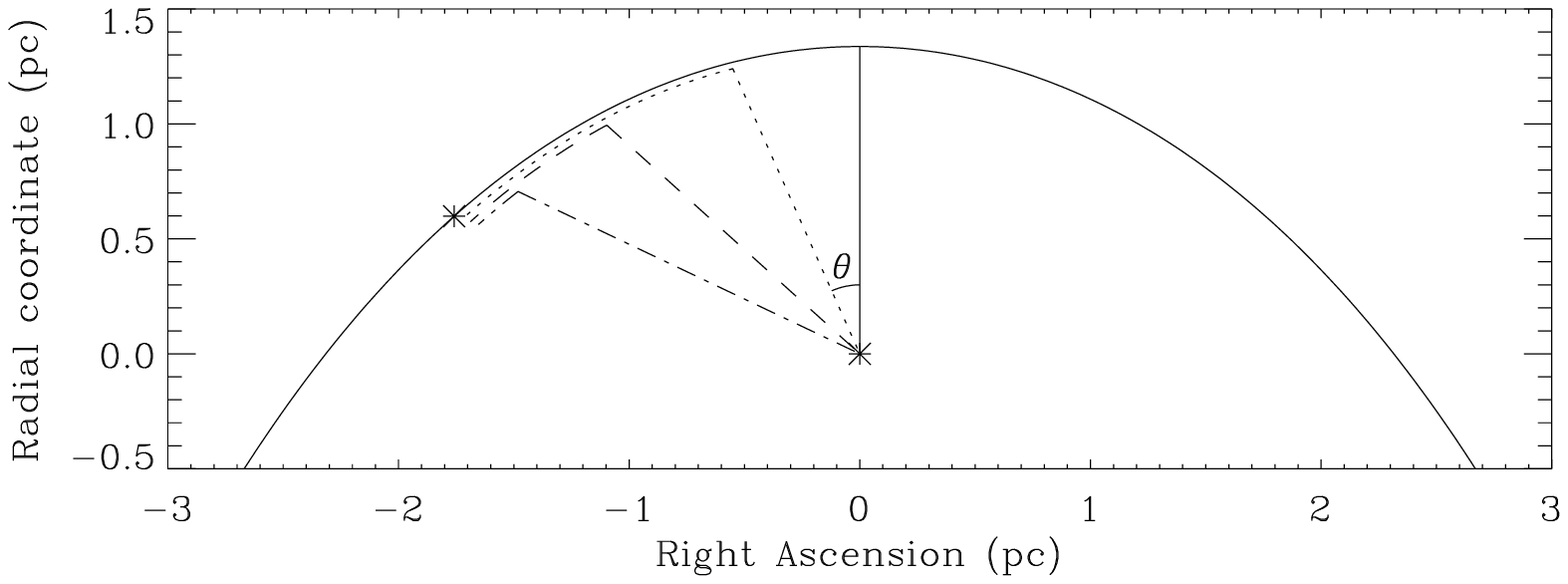}}
\resizebox{\hsize}{!}{\includegraphics{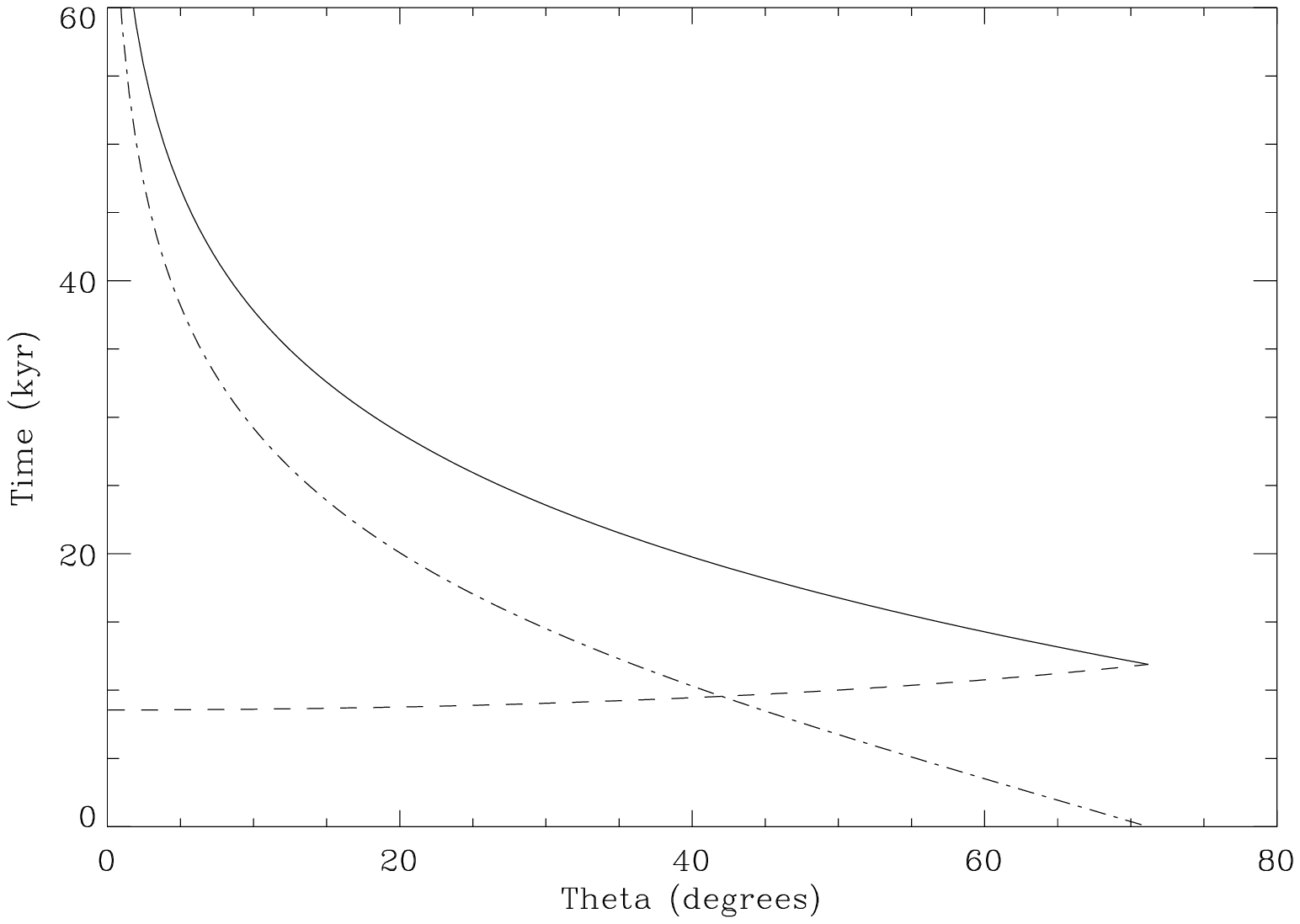}} \caption{
 \emph{Upper panel (a):} Geometry plot, showing the possible paths from the star 
 (asterisk at the origin)
 to an observed structure of gas on the bow shock surface (asterisk at left), 
 characterised by the angle $\theta$.
 \emph{Lower panel (b)}: Travel time for an ejected structure to reach the observed 
 position as a function of $\theta$. The dashed line is the time needed
 to cross the distance from the star to the bow shock.  We assumed an expansion
 velocity of $v_\mathrm{exp} = 150~\mbox{km s}^{-1}$.  The dash-dotted line 
 is the time needed to travel the second part of the path, along the bow shock,
 where the velocity is given by Eq.~(\ref{eq:shockv}).
 The solid line is the total travel time.
 }
 \label{fig:h3857_12}
\end{figure}

The majority of the emission of M1-67 seems to originate from the surface of the bow shock 
in the position-velocity diagram of Fig.~\ref{fig:h3857_10}.  Consider the structure that
we labelled \emph{B} in Fig.~\ref{fig:h3857_11}.  From the Fabry-P\'erot images, we derive its
velocity, as shown in Fig.~\ref{fig:h3857_11}(c). This figure, compared to the first row, fourth 
column image of Fig.~\ref{fig:h3857_10}(a), shows that the structure lies on 
the bow shock surface. We do not know the exact path that it travelled to get to its 
current position, but we do know that this path must be in the plane that is defined by
the main axis of the bow shock and the current position of this structure \emph{B}.  We can
therefore characterise the path by a single angle $\theta$, that is the angle between the
line that joins the front of the bow shock and the star and the direction in which the
structure was expelled, as depicted in Fig.~\ref{fig:h3857_12}(a).  After
the outburst \emph{B}, that caused the structure \emph{B}, occurred, the ejecta first crossed the distance
between the star and the bow shock surface in a straight line and at a constant velocity.
After having reached the bow shock surface, the ejecta collided with it and were eventually 
dragged away along the bow shock surface toward the rear, until they reached their current
position.  The angle $\theta$ can have values ranging from $0\degr$ (expelled exactly toward the
front of the bow shock) to about $70\degr$, the angle at which the structure is observed now, 
which would mean that it moved from the star straight to its current position.
If we assume that the structure \emph{B} was expelled at the same outburst as the freely 
expanding structure \emph{A} discussed above, it should have crossed the distance to the bow 
shock surface with a velocity of $v_\mathrm{exp} = 150~\mbox{km s}^{-1}$.
The velocity on the bow shock, along its surface, is given by Eq.~(\ref{eq:shockv}).  If we then 
integrate over the path, we find a dynamical timescale as a function of $\theta$.  The results 
are displayed in Fig.~\ref{fig:h3857_12}(b).  From this, we find that the dynamical timescale for
structure \emph{B} must be greater than $1.2 \times 10^4$~yr,
for the case where this structure has moved directly from the star to its current position, 
and that it is very likely that the dynamical timescale is in the order of $2 \times 10^4$~yr
(for $\theta \approx 45\degr$).  For this $\theta$, the timescale would even increase to about 
$4 \times 10^4$~yr, if the expansion velocity would be $50~\mbox{km s}^{-1}$.  
For very small $\theta$, the dynamical timescale goes 
to infinity, because the velocity along the bow shock surface drops to zero near the front
(See Fig.~\ref{fig:h3857_05}(b)).

One could debate whether the ejecta can survive the harsh stellar environment for a few times
$10^4$~yr. Indeed, we see that blobs of gas around young, hot stars are photoionised by
the strong radiation field of the star, after which the ionised gas is blown away by the stellar
wind. Being a Wolf-Rayet star, WR~124 clearly has such a strong radiation field. However, the gas
cannot been blown away by the wind, since there is an equilibrium on the surface of the bow shock
between the stellar wind and the ram pressure of the ISM. The gas therefore cannot escape,
other than in the direction toward the back of the bow shock, as we described. In fact, the emission
from M1-67 that we see in Fig.~\ref{fig:h3857_03} is in H$\alpha$, which indicates that the gas
is already ionised. Since we only consider the dynamics of the gas here, it is not important what 
exactly happens to the gas, as long as it does not affect its global motion.

For structure \emph{C} in Fig.~\ref{fig:h3857_11}, roughly the same scenario holds as for structure
\emph{B}. It is very likely to be located on the bow shock surface.  However, this structure
it spread out much more and therefore it is hard to define its exact location.  Furthermore,
it is positioned at the southern cutoff limit of our velocity data. For these reasons, we 
do not discuss its dynamical properties here.  We labelled it in Fig.~\ref{fig:h3857_11} for
completeness, because it is another feature that is clearly visible in the different datasets.

Although there are still large uncertainties in the rough calculations presented here,
it seems very likely that the dynamical timescales of the freely expanding structure \emph{A}
and the shocked structure \emph{B} do not match.  The explanation
that both structures originate from the same bipolar outburst, with an expansion velocity that
is different in different directions seems unlikely, since the directions to these two structures
as seen from the star do not differ too much. This means that 
the two features are likely to originate from different outburst events.  This result strongly suggests
a multiple-outburst scenario, with typical outburst timescales of a few times $10^4$~yr.
This scenario favours an LBV phase rather than an RSG phase that is responsible for the nebula M1-67.

\section{Summary and conclusions}
\label{sec:07}

We tried to fit different freely expanding models to the nebula M1-67
and find that M1-67 is not expanding freely, but that the
outbursts that formed the nebula have been interacting with the ISM.  
Evidence for this is given by the fact that the nebula has a radial 
velocity that is lower than the radial velocity of the star, in 
particular the fact that there is no nebular emission with a
significantly higher radial velocity than that of the star. The 
asymmetry in the amount of emission between the high-velocity 
and low-velocity sides of the nebula and the asymmetry in 
the velocity distribution of the nebula seem to confirm the fact that
there are interactions between the outbursts and the ISM.

We find that the central star WR~124 has a velocity of about
$180~\mbox{km s}^{-1}$ with respect to the surrounding ISM, moving
roughly away from us, which causes a paraboloid-like bow shock
around the star. Because the star is moving away from us, we are 
looking into the hollow bow shock from the rear.  Using the simple
assumption of momentum conservation, we can model the geometry of
the bow shock and the velocity of the gas that is moving along its surface.
This way we can roughly derive the orientation of the bow shock, which
is determined by the spatial velocities of the star and that of the
ISM surrounding it.  The inclination of the main axis of the bow shock 
with respect to the line of sight is about $20\degr$, where the arrow 
that points from the star to the front of the bow shock 
is mainly pointed away from us and slightly to the south.  This small 
inclination is responsible for the fact that we cannot distinguish the 
bow shock in 
the HST image in Fig.~\ref{fig:h3857_03} and that we see the star projected in the centre of the nebula.
The star sits inside this bow shock, at about 1.3~pc from its front.
As input for our bow shock models, we used the observed properties of the stellar wind
($\dot{M}$, $v_\infty$) and assumed an ISM density of $1~m_\mathrm{H}~\mbox{cm}^{-3}$.
These values obviously have quite some uncertainty in them.  We therefore 
once more explicitly mention the rough nature of these calculations. 
The assumptions of a continuous wind and a homogeneous ISM for our models 
are very likely to be incorrect. In fact, an ISM with inhomogeneities could
explain why the observed bow shock does not have a smooth surface.

In general, a run-away star with a wind will form a bow shock. The LBV
eruptions during the late stages of the evolution of massive stars will 
partly collide with the bow shock surface, brighten up and be dragged away 
by the ram pressure of the ISM, along the surface of the bow shock. In 
this way, irregular, or even chaotic nebulae with arcs and rings can be formed by 
intrinsically nicely behaving, possibly even spherical, discrete outbursts.  
The interactions with the ISM make the derivation of dynamical timescales 
for the outbursts much more difficult than in the case of low-velocity stars, 
which blow voids of tens of parsecs in the interstellar medium during their 
O-star phase, so that outbursts occurring in a later stage will indeed expand freely.  However, 
also in the case of a bow shock, a part of each outburst will move toward 
the rear end of this bow shock and remain expanding freely forever.  This part 
might still be fitted in the straightforward way.

For M1-67 we were able to fit the freely expanding part of the outburst \emph{A} in 
Fig.~\ref{fig:h3857_11} to $r \approx 40\arcsec$ and  $v_\mathrm{exp} \approx 150~\mbox{km s}^{-1}$, 
which gives the rough dynamical timescale of $8 \times 10^3$~yr, assuming a distance
of 6.5~kpc.
This dynamical timescale is of the same order as the time it would take this
outburst to cross the distance of $1.3$~pc from the star to the front of the bow shock, 
which means that the far part of this particular outburst might
just have collided with the bow shock.  In fact, the collision of this part of the nebula
is required in order to explain the lack of nebular emission with significantly
higher radial velocities than that of the star.

Part of the nebular emission is found to be on the bow shock surface.  In 
order to get there, this matter could have followed different paths, because
we do not know in which direction the material has been ejected from the star.
We can calculate the amount of time for this matter to have reached its observed
location, as a function of the path.  We did so for the arc that is labelled \emph{B}
in Fig.~\ref{fig:h3857_11}, assuming that the expansion velocity was the same as for 
the freely expanding structure \emph{A}, $v_\mathrm{exp} = 150~\mbox{km s}^{-1}$.
We then find that the age of structure \emph{B} must be greater than $1.2 \times 10^4$~yr, probably
in the order of $2 \times 10^4$~yr, so that this arc most likely originates from a 
different outburst than structure \emph{A}.

The multiple outbursts, the expansion velocity of about $150~\mbox{km s}^{-1}$
and the dynamical timescales in the order of $10^4$~yr are all typical for 
LBVs.  We therefore conclude that an LBV phase preceded the current Wolf-Rayet
phase. The outbursts that occurred during this LBV phase seem to be the most likely 
explanation for the creation of the Wolf-Rayet ring nebula M1-67 around WR~124.

\begin{acknowledgements}
We thank Thierry le F\`evre for his contribution to the freely expanding wind models,
Antonella Nota for the long-slit spectra and Yves Grosdidier for the Fabry-P\'erot images.
\end{acknowledgements}


\bibliographystyle{aa}
\bibliography{h3857}

\end{document}